# Evolution of the pseudogap and excess conductivity of YBa$_2$Cu$_3$O$_{7-\delta}$ single crystals in the course of long-term aging

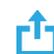

View Online    Export Citation    CrossMark

A. L. Solovjov,[1,2,3,a)] L. V. Bludova,[1] M. V. Shytov,[1] S. N. Kamchatnaya,[2,4] Z. F. Nazyrov,[2,5] and R. V. Vovk[2,4]

AFFILIATIONS

[1]B. Verkin Institute for Low Temperature Physics and Engineering of the National Academy of Sciences of Ukraine, Kharkiv 61103, Ukraine
[2]V. N. Karazin Kharkiv National University, Kharkiv 61022, Ukraine
[3]Institute for Low Temperatures and Structure Research, Polish Academy of Sciences, Wroclaw 50-422, Poland
[4]Ukrainian State University of Railway Transport, Kharkiv 61050, Ukraine
[5]International European University, Kyiv 03187, Ukraine

[a)]Author to whom correspondence should be addressed: solovjov@ilt.kharkov.ua

**ABSTRACT**

The temperature dependences of both fluctuation conductivity (FLC) $\sigma'(T)$ and pseudogap (PG) $\Delta^*(T)$ derived from measurements of resistivity $\rho(T)$ of an optimally doped YBa$_2$Cu$_3$O$_{7-\delta}$ single crystal subjected to long-term storage have been studied. The as-grown sample S1 exhibits characteristics typical of optimally doped YBa$_2$Cu$_3$O$_{7-\delta}$ single crystals containing twins and twin boundaries. Analysis of both FLC and PG showed an unexpected improvement in all characteristics of the sample after 6 years of storage (sample S2), indicating that the effect of twin boundaries is somehow limited. After 17 years of storage, all characteristics of the sample changed dramatically, which indicates a strong influence of internal defects formed during the aging process. For the first time, the temperature dependences of both FLC and PG were obtained after 17 years of storage.

Published under an exclusive license by AIP Publishing. https://doi.org/10.1063/10.0017593

## 1. INTRODUCTION

The discovery of high-temperature superconductors (HTSCs) is undoubtedly one of the landmark events in modern solid state physics. However, despite the efforts of numerous scientific groups and an extraordinary number of publications on HTSCs, the mechanism of superconducting (SC) pairing, which makes it possible to obtain real Cooper pairs at $T \gg 100$ K,[1,2] is still not clear. It is believed that the answer to this and other questions concerning HTSC can be obtained by studying such an unusual phenomenon as the pseudogap (PG), which opens in underdoped cuprates at $T^* \gg T_c$.[3–9] However, the physics of PG is also not completely clear. One of the reasons is that there are two, somewhat opposite, approaches to the interpretation of the pseudogap. Some researchers believe that the appearance of PG is in no way connected with superconductivity.[3,4,9–13] As possible sources of the appearance of PG, not related to superconductivity, are considered superconducting phase fluctuations,[8–10] antiferromagnetic spin fluctuations,[3,11,14] spin density waves (SDW),[12,13,15] charge density waves (CDW),[16,17] leading to the RRW model (reduced d-density wave model),[18] charge ordering (CO),[4,19] and even pair density waves.[20,21]

Another fairly large part of the researchers believe that PG arises as a result of the formation of fermions paired below $T^*$, the so-called local pairs (LPs) or preformed Cooper pairs,[6,10,22–24] and, thus, is a precursor of the transition of HTSCs to the superconducting state.[25,26] Various options for the formation of LPs in HTSCs have been proposed (see Ref. 27 and references therein). We share the idea that LPs arise in the form of so-called strongly bound bosons (SBB),[6,28] whose pairing mechanism is also not completely clear. In a preformed pairs picture, $T_c$ would be controlled by the phase stiffness of the pairs.[10] This idea is confirmed by many experiments[29–32] (and references therein) including results of ARPES,[33–36] tunneling[25,37] and scanning tunneling microscopy[38–40] and supported by corresponding theories[28,41–44] (and references therein). As a result, it turns out that in a wide range of temperatures below $T^*$, the LPs are not true Cooper pairs, but change their properties with decreasing $T$ and already behave like incoherent Cooper pairs near $T_c$.[6,28]

Strictly speaking, it can be argued that, thanks to the efforts of numerous scientific groups, the physics of HTSCs has become clearer to some extent. However, many questions remain regarding the practical application of cuprates. One such issue is the study of the effect of long-term aging of cuprates in air at room temperature.[45–48] As before, the 1–2–3 YBa$_2$Cu$_3$O$_{7-\delta}$ (YBCO) compound is of the greatest interest for a number of reasons. First of all, systems 1–2–3 are the most technologically advanced of all cuprates and do not change the phase composition under the action of various influences.[49,50] In addition, these compounds have a critical temperature $T_c$ of about 90 K at optimal doping, which is above the liquid-nitrogen temperature. The presence of labile oxygen in the 1–2–3 system contributes to the occurrence of structural relaxation processes, which makes it possible to control the electrical transport characteristics of the system. However, the number of papers devoted to the effect of long-term aging of cuprates in air is surprisingly small,[45–48] and the reported experimental data are somewhat contradictory. Most likely this is due to the fact that the data were obtained on samples with different technological prehistory, such as on thin films, ceramics, textured polycrystals and so on.[46,47] At the same time, the disadvantage of YBCO single crystals is the presence of extended planar defects in them, such as twins and twin boundaries (TBs), which must be taken into account.[49,51] The effect of aging on fluctuation conductivity (FLC) and pseudogap could provide additional information about the processes of LPs formation above $T_c$. However, the influence of aging on the FLC and the PG has not yet been studied.

In the paper, we report on the effect of long-term (up to 17 years) aging in air at room temperature on the resistivity, FLC and PG of optimally doped (OD) twinned YBa$_2$Cu$_3$O$_{7-\delta}$ single crystals with SC transition temperature $T_c$ of about 90 K.

## 2. EXPERIMENTAL

The YBa$_2$Cu$_3$O$_{7-\delta}$ single crystals were grown by the solution-melt technique in gold crucible, as described elsewhere.[52] Unless special care are taken,[53] HTSC single crystals always contain twins, which appear during the manufacturing process and make it possible to minimize the internal energy of the crystal.[49,51,52] Rectangular crystals of about $2 \times 0.5 \times 0.02$ mm were selected to perform the resistivity measurements. The smallest parameter of the crystal corresponds to the $c$ axis. A fully computerized setup utilizing the four-point probe technique with stabilized measuring current of up to 10 mA was used to measure the $ab$ plane resistivity, $\rho_{ab}(T)$. Silver epoxy contacts were glued to the extremities of the crystal in order to produce a uniform current distribution in the central region where voltage probes in the form of parallel stripes were placed. Contact resistances below 1 Ω were obtained. Temperatures were measured with a Pt sensor having accuracy about 1 mK. In all cases experimental runs were conducted at rates of about 0.1 K/min near $T_c$ and 0.5 K/min at $T \gg T_c$.

In order to determine the effect of aging, the first measurements of the resistivity $\rho(T) = \rho_{ab}(T)$ were carried out immediately after the separation of crystals from the melt and saturated with oxygen up to an optimal value of $\delta \leq 0.1$ (sample S1). After that, the samples were stored in glass containers and subjected to repeated measurements after 6 (sample S2) and 17 (sample S3) years. Temperature dependencies of resistivity $\rho(T)$ of the initial single crystal with $T_c = 91.66$ K (sample S1) measured at different aging stages are shown in Fig. 1. As expected, S1 has the highest $T_c$ but the lowest resistivity with the lowest slope $d\rho/dT = 0.51$ μΩ cm K$^{-1}$ of the normal-state resistivity above $T^* = 164$ K. After 6 years of storage, the resistivity parameters did not change significantly. Sample S2 has a slightly lower $T_c$, a moderately increased resistivity and a slope $d\rho/dT \approx 0.56$ μΩ cm K$^{-1}$ (Table I), which is in good agreement with the literature data.[54] The more striking result obtained after 17 years of storage. Indeed, the resistivity $\rho(300 K)$ increased by more than 3 times due to the increase in the slope $d\rho/dT = 2.1$ μΩ cm K$^{-1}$ by more than 4 times. However, surprisingly, the SC transition temperature did not change much, as might be expected. Indeed, $T_c = 90.2$ K, if measured by the method shown in the Inset to Fig. 1. Nevertheless, the Inset also shows that the resistive transition of S3 is very wide ($\Delta T_c \approx 7$ K) and indicates the appearance of a second low-temperature phase with $T_c(\rho = 0) \approx 84$ K. Thus, the total $\Delta T_c \approx 16$ K. This form of $\rho(T)$ differs markedly from the "classical" behavior of $\rho(T)$ with a similar slope and $\rho(300 K)$, which is obtained by reducing the charge carrier density $n_f$ in YBCO with a decrease in the oxygen doping level.[54,55] In the latter case, $T_c$ decreases to $\sim 60$ K, but the width of the resistive transition remains small with $\Delta T_c < 4$ K. Thus, we can conclude that the shape of $\rho(T)$ observed after 17 years of storage is due not only to a decrease in $n_f$, which leads to a noticeable increase in resistivity, but also indicates the occurrence of significant structural distortions in the crystal. This conclusion is confirmed by the results of the analysis of the fluctuation conductivity and especially the pseudogap.

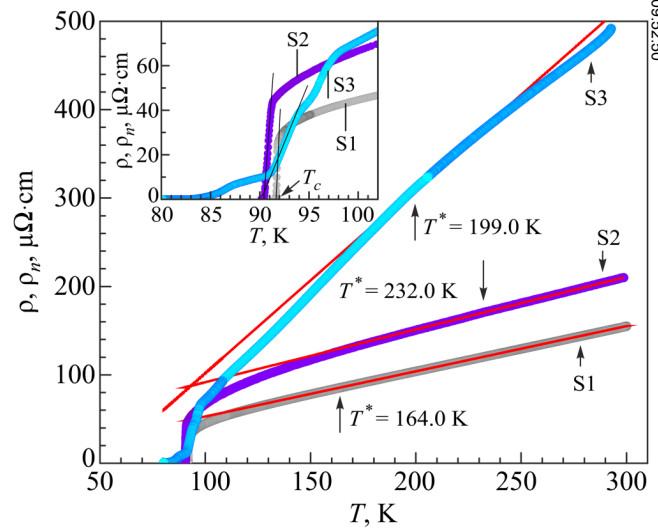

**FIG. 1.** Temperature dependences of the resistivity of an optimally doped YBa$_2$Cu$_3$O$_{7-\delta}$ single crystal ($T_c \approx 91.7$ K) at different stages of aging. Samples: as-grown (S1), after 6 years (S2), and after 17 years of storage (S3). The arrows designate $T^*$. Inset shows the resistive transitions to the superconducting state for all samples, which determine $T_c$.

TABLE I. Resistive and FLC parameters of YBa$_2$Cu$_3$O$_{7-\delta}$ single crystal at different stages of aging.

| Sample, years | $\rho$(300 K), $\mu\Omega$ cm | $\rho$(100 K), $\mu\Omega$ cm | $T_c$, K | $T_c^{mf}$, K | $T_G$, K | $T_0$, K | $T_{01}$, K | $C_{3D}$ | $d_{01}$, Å | $\xi_c(0)$, Å |
|---|---|---|---|---|---|---|---|---|---|---|
| S1(0)  | 155   | 44.87 | 91.66 | 91.72 | 91.7 | 91.8 | 92.4  | 0.3  | 4.1 | 0.36 |
| S2(6)  | 209.8 | 67.2  | 90.5  | 90.77 | 90.9 | 91.4 | 94.5  | 1.05 | 4.2 | 0.87 |
| S3(17) | 491.4 | 73.22 | 90.2  | 95.3  | 96.1 | 98.0 | 101.7 | 2.1  | 6.7 | 2.0  |

## 3. RESULTS AND DISCUSSION

### 3.1. Fluctuation conductivity

The deviation of $\rho(T)$ from linearity below the PG temperature $T^*$ (Fig. 1) leads to the appearance of excess conductivity $\sigma'(T)$ determined by a simple formula

$$\sigma'(T) = \sigma(T) - \sigma_N(T) = [1/\rho(T)] - [1/\rho_N(T)]. \quad (1)$$

In Eq. (1), $\rho(T) = \rho_{ab}(T)$ and $\rho_N(T) = \rho_0 + aT$ determines the resistivity of a sample in the normal state extrapolated toward low temperatures, where $\rho_0$ is a residual resistance and $a = d\rho/dT$. This way of determination of $\rho_N(T)$, which is widely used for calculation $\sigma'(T)$ in cuprates[32] (and references therein), has found validation in the nearly antiferromagnetic Fermi liquid (NAFL) model.[13] As usually, to determine $T^*$ more precisely the criterium $[\rho(T) - \rho_0] = aT^2$ was used. The superconducting transition temperature $T_c$ is determined by extrapolation of the linear part of the resistive transition to $\rho(T) = 0$ (Inset in Fig. 1). Here we focus on the analysis of FLC and PG derived from measured excess conductivity within the local pair (LP) model.[6,28,55] The determined from the analysis samples parameters are listed in Tables I and II.

The first step to begin the analysis is the determination of mean field critical temperature $T_c^{mf}$.[32,55] Here, $T_c^{mf} > T_c$ is the critical temperature in the mean-field approximation, which separates the FLC region from the region of critical fluctuations or fluctuations of the SC order parameter $\Delta$ directly near $T_c$ (where $\Delta < k_BT$), neglected in the Ginzburg–Landau (GL) theory.[56,57] In the LP model, it is believed that PG and excess conductivity are due to the formation of local pairs below $T^*$.[6,28,32,55] The properties of the LPs are determined by the coherence length along the $c$ axis

$$\xi_c(T) = \xi_c(0)\left[(T - T_c^{mf})/T_c^{mf}\right]^{-1/2} = \xi_c(0)\,\varepsilon^{-1/2}$$

TABLE II. Pseudogap parameters of YBa$_2$Cu$_3$O$_{7-\delta}$ single crystal at different stages of aging

| Sample, years | $T^*$, K | $\alpha$ | $\varepsilon_{c0}^*$ | $A_4$ | $D^*$ | $\Delta^*T_G$, K | $T_{\text{pair}}$, K | $\Delta^*T_{\text{pair}}$, K |
|---|---|---|---|---|---|---|---|---|
| S1(0)  | 164 | 6.7 | 0.15 | 8.4 | 5.0 | 229.3 | 115.2 | 202.5 |
| S2(6)  | 232 | 4   | 0.25 | 23  | 5.0 | 225.2 | 143.0 | 266.2 |
| S3(17) | 199 | 3.4 | 0.29 | 41  | 5.0 | 225.8 | 163.3 | 249.4 |

Refs. 58 and 59, where

$$\varepsilon = (T - T_c^{mf})/T_c^{mf} \quad (2)$$

is the reduced temperature. Thus, the correct determination of $T_c^{mf}$ is decisive in FLC and PG analysis.

It was convincingly shown that FLC measured for all without exception HTSC's always demonstrates a transition from 2D ($\xi_c(T) < d$) into 3D ($\xi_c(T) > d$) state, as $T$ draws near $T_c$, where $d \approx 11.68$ Å is the unit cell size along the $c$ axis.[60] This is most likely a consequence of Gaussian fluctuations of the order parameter in 2D metals,[61–63] which prevent any phase coherency organization in 2D compounds. As a result, the critical temperature of an ideal 2D metal is found to be zero (Mermin–Wagner–Hoenberg theorem), and a finite value is obtained only when three-dimensional effects are taken into account.[61–64] That is why, the FLC in the 3D state is always extrapolated by the standard 3D equation of the Aslamazov–Larkin (AL) theory, which determines the FLC in any 3D system:[65]

$$\sigma'_{AL3D} = C_{3D}\frac{e^2}{32h\xi_c(0)}\varepsilon^{-1/2}, \quad (3)$$

where $C_{3D}$ is a scaling factor.

This means that the conventional 3D FLC is realized in HTSC's as $T \to T_c$.[6,32,59] From Eq. (3), one can easily obtain $\sigma'^{-2} \sim \varepsilon \sim (T-T_c^{mf})/T_c^{mf}$. Obviously, $\sigma'^{-2} = 0$, when $T = T_c^{mf}$ as shown in Fig. 2 on the example of sample S1. This way of $T_c^{mf}$ determination was proposed by Beasley[57] and substantiated in various FLC experiments.[6,32,53,55] Apart from $T_c^{mf} = 91.72$ K and $T_c$, the figure shows the Ginzburg temperature $T_G$, down to which fluctuation theories work as $T$ decreases, and the 3D-2D crossover temperature $T_0$, which limits the region of 3D-AL fluctuations from above.

Next, having determined $T_c^{mf}$, we plotted ln $\sigma'$ vs ln $\varepsilon$ for sample S1, S2, and S3, as shown in Fig. 3. Near $T_c$, the FLC is perfectly approximated by the fluctuation contribution of the AL theory for 3D systems [Eq. (3)]. In double logarithmic coordinates, these are red straight lines (1) with a slope $\lambda = -1/2$. This confirms the above conclusion that the classical fluctuation mode of 3D-AL is always realized in HTSCs near $T_c$,[6,32,53,66] since here $\xi_c(T) \gg d$, and fluctuation Cooper pairs (FCPs) can interact throughout the sample volume.

Above $T_0$ (ln $\varepsilon_0$ in the figure), where $\xi_c(T) < d$, the three-dimensional regime ends. But, it is still $\xi_c(T) > d_{01} \approx 4$ Å, which is the distance between conducting planes CuO$_2$,[60] and $\xi_c(T)$ still connects the CuO$_2$ planes with the Josephson interaction. This is a 2D fluctuation mode, which is realized in the interval from $T_0$ to $T_{01} \approx 92.4$ K ($\ln \varepsilon_{01}$ in the figure) (S1) and is described by the Maki–Thompson (MT) term of the Hikami–Larkin (HL) theory:[58]

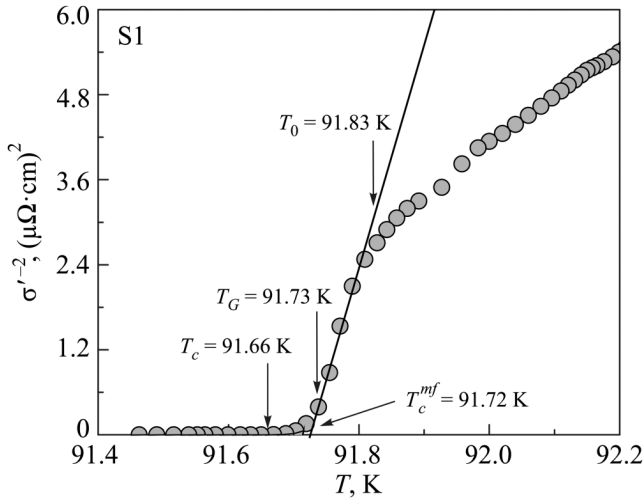

**FIG. 2.** Temperature dependence of the inverse square of the excess conductivity, $\sigma'^{-2}(T)$, for sample S1 (gray dots), which determines $T_c^{mf} = 91.72$ K. The arrows also indicate $T_c$, the Ginzburg temperature $T_G$, and the 3D-2D crossover temperature $T_0$.

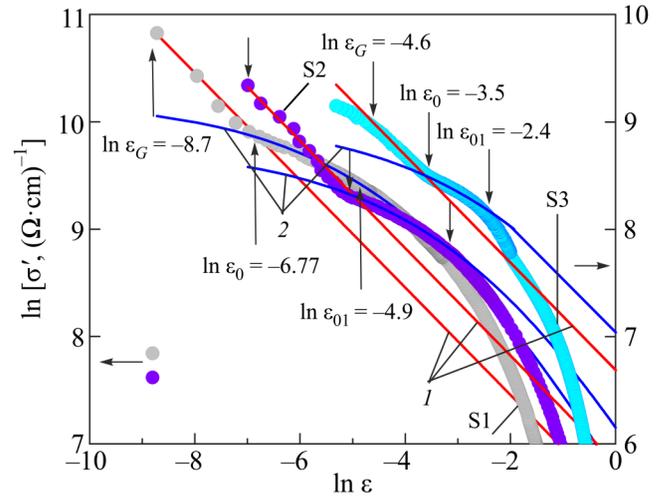

**FIG. 3.** Dependences $\ln \sigma'$ vs $\ln \varepsilon$ for sample S1, S2, and S3 in the range of SC fluctuations near $T_c$ in comparison with 3D-AL (red straight lines *1*) and 2D-MT (blue curves *2*) fluctuation theories. All characteristic temperatures $T_G$, $T_0$, and $T_{01}$ are marked with arrows. The arrows for S2 are unlabeled, and the Y axis for S3 has been shifted to make the drawing more readable.

$$\sigma'_{MT} = C_{2D} \frac{e^2}{8d\hbar} \frac{1}{1-\alpha/\delta} \ln\left((\delta/\alpha)\frac{1+\alpha+\sqrt{1+2\alpha}}{1+\delta+\sqrt{1+2\delta}}\right)\varepsilon^{-1}. \quad (4)$$

These are the blue curves (*2*) in the figure. Accordingly, in Eq. (4), $\alpha = 2[\xi_c(0)/d]^2 \varepsilon^{-1}$ is a coupling parameter,

$$\delta = \beta \frac{16}{\pi\hbar} \left[\frac{\xi_c(0)}{d}\right]^2 k_B T \tau_\varphi \quad (5)$$

is the pair-breaking parameter, and $\tau_\varphi$ is the phase relaxation time

$$\tau_\varphi \beta T = \pi\hbar/8k_B\varepsilon_0 = A/\varepsilon_0, \quad (6)$$

where $A = 2.998 \cdot 10^{-12}$ s K. The factor $\beta = 1.203(l/\xi_{ab})$, where $l$ is the mean-free path and $\xi_{ab}(0)$ is the coherence length in the $ab$ plane considering the clean limit approach $l > \xi$, which is always takes place in HTSCs.[5,6,32,66]

Thus, at $T = T_0$, a 3D-2D (AL-MT) crossover occurs. Obviously, at $T_0$ $\xi_c(T_0) = \xi_c(0)\varepsilon_0^{-1/2} = d$, which makes it possible to determine the coherence length along the $c$ axis

$$\xi_c(0) = d\sqrt{\varepsilon_0}. \quad (7)$$

Having determined $T_0$ and $\ln \varepsilon_0$, by Eq. (7) we find $\xi_c(0) = (0.36 \pm 0.01)$ Å (S1) (Table I), which is a typical value of $\xi_c(0)$ for OD YBCO single crystals with twins and the same $T_c = 91.7$ K.[67] Accordingly, $\xi_c(T_{01}) = \xi_c(0)\varepsilon_{01}^{-1/2} = d_{01}$ which makes it possible to estimate the distance between the conducting CuO$_2$ planes.[32,53,67] To find $d_{01}$, we use the experimental fact that $\xi_c(0)$ is already determined by the crossover temperature $T_0$ [Eq. (7)]. Hence, the condition $\xi_c(0) = d\sqrt{\varepsilon_0} = d_{01}\sqrt{\varepsilon_{01}} = (0.36 \pm 0.01)$ Å (S1) is satisfied. Since $d = c = 11.68$ Å we get: $d_{01} = d(\sqrt{\varepsilon_0}/\sqrt{\varepsilon_{01}}) = (4.14 \pm 0.05)$ Å for S1, which, in fact, is the interplanar distance in YBCO.[60] In the same way, the corresponding values of $d_{01}$ were found for all other samples (Table I). From Fig. 3 the following evolution of the FLC is traced. S1 (gray dots) shows the dependence of $\ln \sigma'$ vs $\ln \varepsilon$, which is typical for OD single crystals containing TBs with very pronounced enhanced 2D-MT fluctuations. Here, the enhanced fluctuation means that the deviation of the 2D-MT fluctuations above the extrapolated 3D-AL red line, which we usually designate as $\Delta \ln \sigma'$,[32,49] is quite large. Surprisingly, after 6 years of storage (purple dots) S2 demonstrate the FLC behavior characteristic of Y1–2–3 systems with a reduced number of defects, namely, the range of 3D fluctuations noticeably increased, while $\Delta \ln \sigma'$, on the contrary, decreased, but the absolute value $\ln \sigma'$ practically did not change. Interestingly, S2 has $C_{3D} \approx 1$ (Table I), which is typical value for the well-structured YBCO thin films.[55] In turn, as expected, S3 (blue dots) has the smallest absolute FLC value. Strictly speaking, the experimental curve looks somewhat distorted, and 2D-MT fluctuations are almost completely suppressed. Such $\ln \sigma'$ vs $\ln \varepsilon$ with largest $C_{3D} \approx 2.1$ (Table I) is a hallmark of the behavior of FLC in Y1–2–3 systems with defects.[68] Our calculations also indicate an increase in $d_{01}$ by more than 1.6 times in S3 (Table I). These data allow us to conclude that S3 has an increased number of defects, which is consistent with the results of resistivity measurements. It would be interesting to find confirmation of these results in the behavior of PG.

### 3.2. Pseudogap analysis

The pseudogap $\Delta^*(T)$ is the suppression of density of states at the Fermi level with energy $|\varepsilon| \leq \Delta_{PG}$ at temperature $T < T^*$, where $T^*$ is markedly higher than $T_c$, as mentioned above. In resistivity measurements, PG is manifested as a deviation of resistivity from linear behavior at temperatures below $T^*$ in underdoped cuprates. To get information about PG from $\sigma'(T)$, a proper relationship is needed, which would determine all experimental data in the entire range from $T^*$ to $T_c$ and explicitly include the PG parameter $\Delta^*(T)$. According to the local-pair model, $\Delta^*(T)$ can be correlated to the excess conductivity $\sigma'(T)$ according to the formula:[6,55]

$$\sigma'(T) = A_4 \frac{e^2 \left(1 - \frac{T}{T^*}\right) \exp\left(\frac{-\Delta^*}{T}\right)}{16\hbar\xi_c(0)\sqrt{2\varepsilon_{c0}^*} \sinh\left(2\frac{\varepsilon}{\varepsilon_{c0}^*}\right)}. \quad (8)$$

Here, $\varepsilon_{c0}^*$ is a parameter of the theory,[69] which can be determined from the experimental data, $A_4$ is a fitting coefficient. $(1 - T/T^*)$ denotes the number of LPs formed at $T \leq T^*$, and $\exp(-\Delta^*/T)$ describes the dynamics of the destruction of LPs as $T$ approaches $T_c$. For a direct determination of $\Delta^*(T)$, Eq. (8) can be re-written as follows:

$$\Delta^*(T) = T \ln\left[A_4\left(1 - \frac{T}{T^*}\right)\frac{1}{\sigma'(\varepsilon)}\frac{e^2}{16\hbar\xi_c(0)}\frac{1}{\sqrt{2\varepsilon_{c0}^*}\sinh\left(2\frac{\varepsilon}{\varepsilon_{c0}^*}\right)}\right], \quad (9)$$

where $\sigma'(\varepsilon)$ is the experimentally measured excess conductivity over the whole temperature interval from $T^*$ down to $T_G$. As can be seen from Figs. 4(a) and 4(b), Eq. (8) (indicated by the solid red curves in the figures) fits reasonably with the experimental data of both S1 (a), S2 (not shown), and S3 (b) samples.

This indicates the validity of LP model adopted in this work. The fitting procedure yields to the determination of important parameters such as the theoretical parameter $\varepsilon_{c0}^*$, the numerical factor $A_4$, and the pseudogap $\Delta^* = \Delta^*(T_G)$. Indeed, there is an exponential dependence of $\sigma'(T)$ vs $\varepsilon$, more precisely $[1/\sigma'(T) \exp(\varepsilon)]^{32,53,55,69}$ in the temperature range $\ln \varepsilon_{01} < \ln \varepsilon < \ln \varepsilon_{02}$ [Fig. 4(b)] or correspondingly $\varepsilon_{c01} < \varepsilon < \varepsilon_{c02}$ (Inset to Fig. 4(b), S3). Such exponential dependence has been shown to be a typical characteristic of HTSCs.[6,55,69]

Obviously, in coordinates $\ln(1/\sigma')$ vs $\varepsilon$, this is a straight line with a slope $\alpha^* = 1/\varepsilon_{c0}^*$. The Inset of Fig. 4(b) shows a typical example of linear dependence of $\ln(1/\sigma')$ vs $\varepsilon$ with a slope of $\alpha^* = 3.4$ obtained for sample S3, which allows the determination of $\varepsilon_{c0}^* = 1/\alpha^* = 0.29$. Performing similar calculations for S1 and S2 we get results listed in Table II. The same linear tendency has been also observed in the case of YBCO-PrBCO with a slope $\alpha^*=1.7$,[32] $HoBa_2Cu_3O_{7-y}$ single crystals with a slope $\alpha^* =1.24$[51] and many others.[6,53,55]

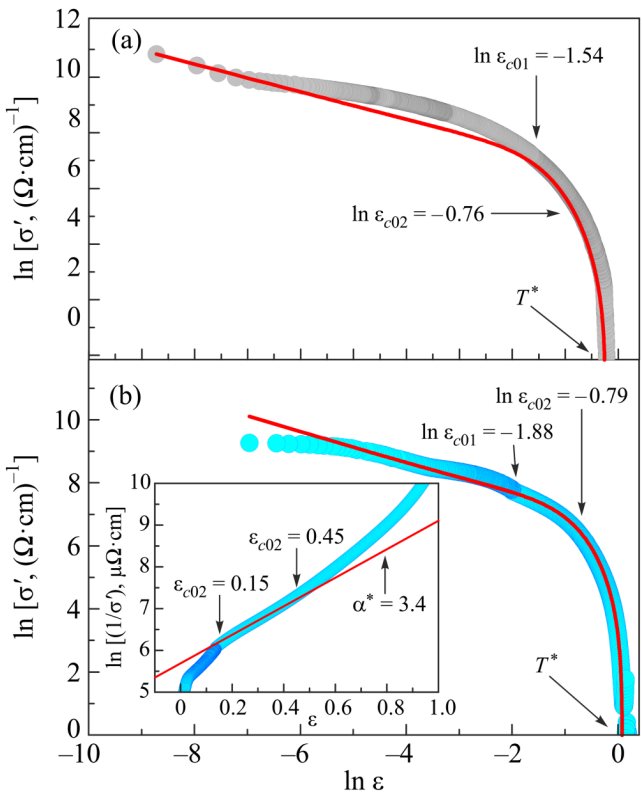

FIG. 4. Dependences of $\ln \sigma'$ vs $\ln \varepsilon$ of the OD YBa$_2$Cu$_3$O$_{7-\delta}$ single crystal in the entire temperature range from $T^*$ down to $T_G$ at different stages of aging. S1 (gray dots) as-grown and S3 (blue dots) after 17 years of storage. Red curves are approximation of experimental data by Eq. (8) with a set of parameters given in the text. Inset: $\ln(1/\sigma')$ as a function of $\varepsilon$ for S3 (b). The straight red line denotes the linear part of the curve between $\varepsilon_{c01} = 0.15$ and $\varepsilon_{c02} = 0.45$. The corresponding values of $\ln(\varepsilon_{c01})$ and $\ln(\varepsilon_{c02})$ are indicated by arrows on the main panels for both samples. The slope $\alpha^* = 3.4$ determines the parameter $\varepsilon_{c0}^* = 1/\alpha^* = 0.29$ for S3.

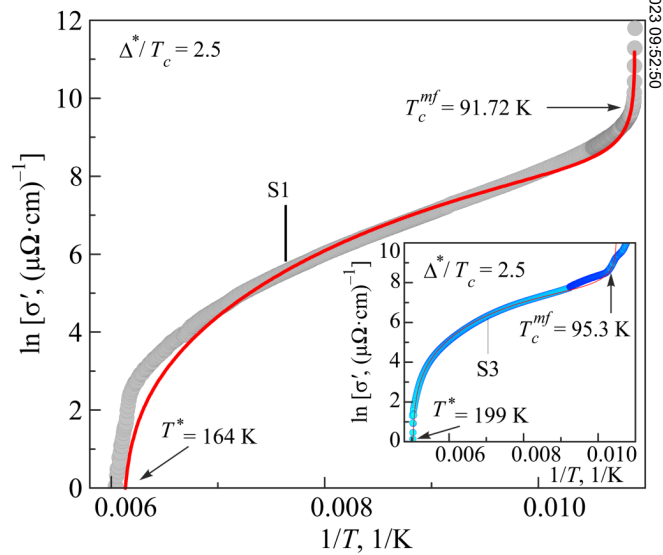

FIG. 5. $\ln \sigma'$ as a function of $1/T$ for samples S1 (gray dots) and S3 (blue dots, Inset) in the entire temperature range from $T^*$ down to $T_G$. Red curves are approximation of experimental data by Eq. (8) with a set of parameters given in the text. The best approximation is achieved at the value of the BCS ratio $D^* = 2\Delta^*(T_G)/k_B T_c = 5$ for both samples.

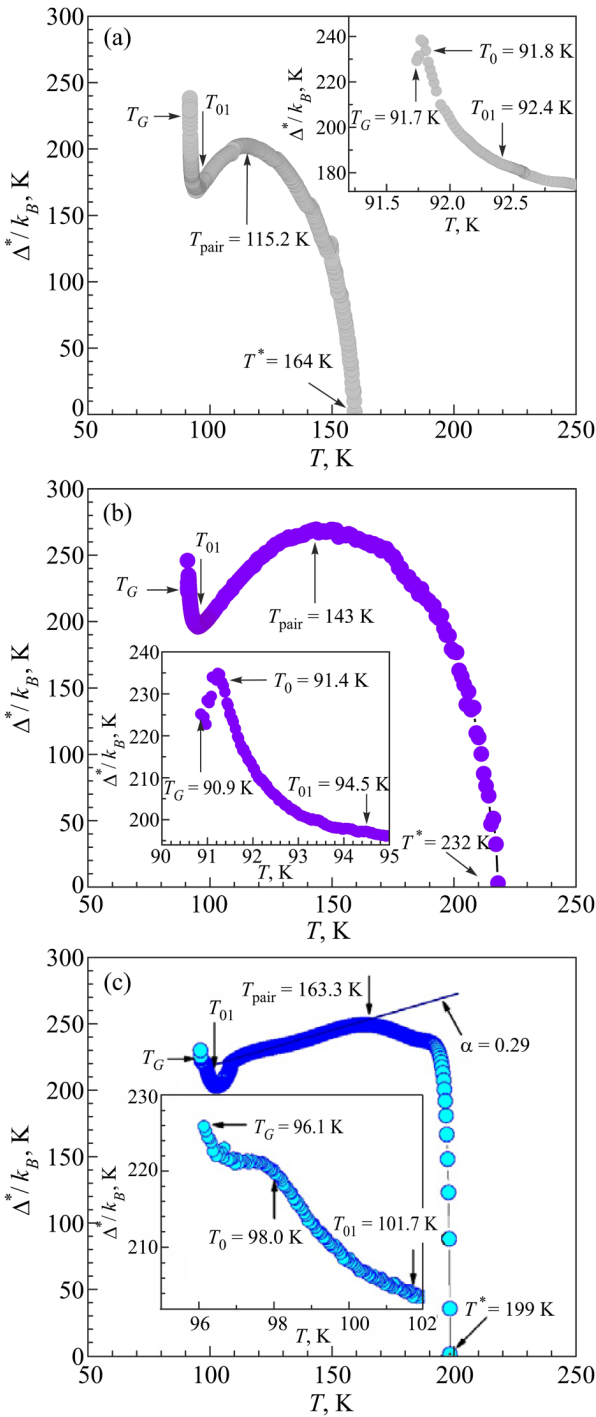

**FIG. 6.** Temperature dependences of the pseudogap $\Delta^*(T)$ for the as-grown OD single crystal of $YBa_2Cu_3O_{7-\delta}$ sample S1 (a), after 6 years of storage (sample S2) (b), after 17 years of storage (sample S3) (c). Each Inset shows $\Delta^*(T)$ in the range from $T_G$ to $T_{01}$. The arrows indicate the corresponding characteristic temperatures.

To proceed with the PG analysis, we also need to define the explicit $\Delta^*(T_G)$ in Eq. (8). It was clearly established that $\Delta^*(T_G) = \Delta(0)$, where $\Delta(0)$ is the SC order parameter at $T = 0$,[37,70] which makes it possible to determine the BCS ratio from the measurements of the PG.[6,32,53]

Figure 5 depicts the variations of $\ln \sigma'(T)$ as a function of $1/T$ for both S1 and S3 samples, which turned out to be ultra-sensitive to the $\Delta^*(T_G)$ value.[32,55] From this figure, the value of $\Delta^*(T_G)$ can be estimated by fitting the data according to Eq. (8). The best fit was obtained when $\Delta^*(T_G)/k_B T_c \approx 2.5$ for both S1 and S3 samples (solid red curves in the figure). Hence, the BCS ratio $D^* = 2\Delta(0)/k_B T_c = 2\Delta^*(T_G)/k_B T_c \sim 5$ for all samples (Table II). Note that $D^* = (5 \pm 0.2)$ is a typical value for YBCO, which implies a strong coupling limit for $d$ wave superconductors.[6,32]

Having determined both $\varepsilon_{c0}^*$ and $\Delta^*(T_G)$ and using the previously determined $T^*$ and $\xi_c(0)$, the $A_4$ scale factor can now be estimated. To do this, we describe the $\ln \sigma'$ versus $\ln \varepsilon$ data by Eq. (8) with a set of found parameters and use $A_4$ as a fitting parameter (Fig. 4). The results of approximation are presented by solid red curves in the main panels of Fig. 4 for both S1 (a) and S3 (b). The deviation of theory from the experiment in the region of 2D fluctuations (S1) is associated with the revealed enhanced 2D-MT fluctuations (Fig. 3), which were not taken into account in the LP model.[32,55] The obtained values of A4 are listed in Table II. Now we have a complete set of parameters to plot $\Delta^*(T)$ for all the samples studied. Figure 6 shows the dependences of $\Delta^*(T)$ for samples S1 [(a), gray dots], S2 [(b), purple dots], and S3 [(c), blue dots]. $\Delta^*(T)$ was determined using Eq. (9) by considering the parameters' values $\sigma'(T)$, $T^*$, $T_c^{mf}$, $\varepsilon_{c0}^*$, and $\xi_c(0)$ extracted from experiments and the numerical parameter $A_4$ deduced from fitting procedures (refer to Tables I and II). The curves plotted in Fig. 6 show a specific shape with maxima and minima at particular temperatures. Figure 6(a) (gray dots) displays $\Delta^*(T)$ for S1 calculated using Eq. (9) with the following set of parameters derived from the experiment within the LP model: $T^* = 164$ K, $T_c^{mf} = 91.72$ K, $\xi_c(0) = 0.36$ Å, $\varepsilon_{c0}^* =0.15$, $A_4 = 8.4$. As expected, the resulting form of $\Delta^*(T)$ with a relatively narrow and low maximum at $T_{pair} \approx 115$ K and a dip minimum at $T_{01} = 92.4$ K followed by a sharp increase of $\Delta^*(T)$ up to $\Delta^*(T_G) \approx 229$ K in the very narrow range of SC fluctuation (see Inset) is typical of a optimally doped HTSC single crystals, containing tweens and TBs[49] (and references therein). This sharp $\Delta^*(T)$ enhancement can be associated with a sudden increase of the superfluid density, $n_s$, that is the density of fluctuating Cooper pairs (short-range phase correlations) forming in the sample when $T$ approaches $T_c$. The Inset shows details of the low-temperature behavior of $\Delta^*(T)$ in S1 with minimum at $T_{01}$, maximum at about $T_0$ and final minimum at $T_G$, which is also characteristic of all HTSCs (see Fig. 12 in Ref. 32). The range of SC fluctuations $\Delta T_{fl} = T_{01} - T_G \approx 0.7$ K is very small but comparable with $\Delta T_{fl}$ obtained for OD YBCO single crystals with TBs.[67]

Figure 6(b) (purple dots) displays $\Delta^*(T)$ for S2 calculated by Eq. (9) with the corresponding set of parameters listed in Tables I and II. As expected, the now obtained form of $\Delta^*(T)$ with a broad maximum at $T_{pair} \approx 143$ K and a minimum at $T_{01} = 94.5$ K followed by a moderate increase in $\Delta^*(T)$ up to $\Delta^*(T_G) \approx 225$ K is typical of well-structured HTSCs, both thin films[55] and single crystals[53] (and references therein). Inset shows the low-temperature behavior of $\Delta^*(T)$ in S1 with minimum at $T_{01}$, maximum at about $T_0$ and final minimum at $T_G$, which is characteristic of all HTSCs, as mentioned above. The range of SC fluctuations has also increased up to $\Delta T_{fl} = T_{01} - T_G \approx 3.6$ K which is comparable with $\Delta T_{fl}$ obtained for OD YBCO untwined single crystals.[53] This result confirms the somewhat unexpected conclusion of the FLC study that the crystal structure of the sample somehow improved after 6 years of storage.

Finally, Fig. 6(c) (blue dots) shows $\Delta^*(T)$ for S3 also calculated by Eq. (9) with the corresponding set of parameters listed in Tables I and II. A rather peculiar $\Delta^*(T)$ was found in this case, which does not find a direct analogue in our data bank $\Delta^*(T)$. The slope of the $\Delta^*(T)$ changes sharply several times, indicating a significant change in the LPs interaction mechanism in the sample, most likely caused by defects. The linear part of $\Delta^*(T)$, which ranges from $T_{pair}$ down to about $T_{01}$ is characteristic of OD YBCO single crystals with defects produced, for example, by high-energy electron irradiation.[49] The range of SC fluctuations $\Delta T_{fl} = T_{01} - T_G \approx 5.6$ K is relatively large, but the shape of $\Delta^*(T)$ near $T_c$ is completely distorted by defects (see Inset). Ultimately, we can conclude that it is likely that the $\Delta^*(T)$ revealed in our experiment after 17 years of storage of the YBCO single crystal is formed by defects and can be considered as a typical form of $\Delta^*(T)$ after long-term storage.

## 4. CONCLUSION

The study of the effect of long-term aging on fluctuation conductivity $\sigma'(T)$ and pseudogap $\Delta^*(T)$ in optimally doped $YBa_2Cu_3O_{7-\delta}$ sheds new light on the inter-particle interaction in cuprates. For the first time, three types of temperature dependences of FLC and PG were observed depending on the aging time. As expected, the as-grown sample S1 exhibits the highest $T_c$ and the lowest resistivity $\rho(T)$. Analysis of the $\sigma'(T)$ and $\Delta^*(T)$ derived from resistivity measurements within the local pair model showed that S1 has characteristics typical of OD $YBa_2Cu_3O_{7-\delta}$ single crystals containing twins and twin boundaries. However, very surprisingly, FLC and PG analysis performed after 6 years of storage showed an unexpected improvement in all sample characteristics (S2). We believe this demonstrates that $YBa_2Cu_3O_{7-\delta}$ undergoes some specific rearrangement of its crystal structure with time, which somehow limits the effect of TBs. As a result $\sigma'(T)$ and $\Delta^*(T)$ obtained for S2 are almost the same as those usually observed for well-structured YBCO.

In contrast to these results, the values of $\sigma'(T)$ and $\Delta^*(T)$ obtained after 17 years of storage (sample S3) changed dramatically. Resistivity $\rho(300 K)$, $\rho(100 K)$, and their linear slope $a = d\rho/dT$ increased by more than 3, 1.6, and 4 times, respectively. The resistive transition of S3 became very wide ($\Delta T_c \approx 7$ K) and pointed out the appearance of a second low-temperature phase with $T_c(\rho=0) \approx 84$ K. Thus, the total $\Delta T_c \approx 16$ K. This form of $\rho(T)$ differs markedly from the "classical" behavior of $\rho(T)$ with a similar slope and $\rho(300 K)$, which is obtained by reducing the charge carrier density $n_f$ in YBCO with a decrease in the oxygen doping level.[54,55] Correspondingly, the shape of both $\sigma'(T)$ and $\Delta^*(T)$ has also changed greatly. It was found that the fluctuation contribution of 2D-MT is almost completely suppressed. In addition, S3 exhibits the largest scaling factor $C_{3D} = 2.1$ and an unexpectedly increased distance between conducting $CuO_2$ planes $d_{01} = 6.7$ Å, which is about 1.6 times greater than that estimated for YBCO.[60] All these data indicate the presence of a large number of defects leading to significant structural distortions in the crystal. This conclusion is confirmed by the results of the analysis of the pseudogap. A rather peculiar $\Delta^*(T)$ was found in this case, which does not find a direct analogue in our data bank $\Delta^*(T)$. The slope of the $\Delta^*(T)$ changes sharply several times, indicating a significant change in the LPs interaction mechanism in the sample, most likely caused by defects formed during the aging process. The range of SC fluctuations $\Delta T_{fl} = T_{01} - T_G \approx 5.6$ K is relatively large, but the shape of $\Delta^*(T)$ near $T_c$ is completely distorted by defects. Ultimately, we can conclude that it is likely that the $\Delta^*(T)$ revealed in our experiment after 17 years of storage of the YBCO single crystal is formed by defects and can be considered as a typical form of $\Delta^*(T)$ after long-term storage.

## ACKNOWLEDGMENTS

We acknowledge support from the European Federation of Academies of Sciences and Humanities though GrantALLEA (A.L.S.). A.L.S. also thanks the Division of Low Temperatures and Superconductivity, INTiBS Wroclaw, Poland, for their hospitality.